\documentclass[doublecol]{epl2}
\usepackage{graphicx}
\usepackage{amsmath} 
\usepackage{amssymb} 
\usepackage{siunitx}

\newcommand{\bv}{{\mathbf v}}
\newcommand{\bV}{{\mathbf V}}

\newcommand{\hatn}{\hat{n}}

\newcommand{\gat}{\gamma_a}
\newcommand{\dat}{\Delta}

\newcommand{\ri}{R_I}
\newcommand{\eps}{\epsilon}
\newcommand{\tmax}{\text{max}}

\title{Granular Brownian motion with dry friction}

\author{A. Gnoli$^1$, A. Puglisi$^1$, and H. Touchette$^2$}
\shortauthor{A. Gnoli, A. Puglisi, and H. Touchette}
\institute{$^1$ CNR-ISC and Dipartimento di Fisica, Universit\`a
Sapienza - p.le
A. Moro 2, 00185, Roma, Italy\\$^2$ School of Mathematical Sciences,
Queen Mary
University of London, London E1 4NS, UK}

\abstract{The interplay between Coulomb friction and random
  excitations is studied experimentally by means of a rotating probe
  in contact with a stationary granular gas. The granular material is
  independently fluidized by a vertical shaker, acting as a ``heat
  bath'' for the Brownian-like motion of the probe. Two ball
  bearings supporting the probe exert nonlinear Coulomb friction upon
  it. The experimental velocity distribution of the probe,
  autocorrelation function, and power spectra are compared with the
  predictions of a linear Boltzmann equation with friction, which is
  known to simplify in two opposite limits: at high collision
  frequency, it is mapped to a Fokker-Planck equation with nonlinear
  friction, whereas at low collision frequency, it is described by a
  sequence of independent random kicks followed by friction-induced
  relaxations. Comparison between theory and experiment in these two
  limits shows good agreement. Deviations are observed at very small
  velocities, where the real bearings are not well modeled by Coulomb
  friction.}


\begin{document}

\maketitle

\section{Introduction} 

The role of fluctuations in solid-solid interactions with friction has
been increasingly studied in the last years for different systems and
different scales. Starting from the 60s, engineers  studied the effect of
noise on dry friction and dry contacts as a way to model the
stability of buildings under earthquakes; see,
e.g.,~\cite{CD61,AC68,AC68b}. More recently, physicists have started
studying similar problems, but from a more microscopic point of view,
by looking at the effects of noise on ``small'' systems with few
degrees of freedom, such as those studied, for example, in
nanofriction experiments~\cite{RG04}, particle separation \cite{EG06},
ratchets and granular motors \cite{BBG06,FAU07,TWV11}, as well as
droplet dynamics on surfaces~\cite{DCG05,MC10,GC10}, which involves
forces similar to dry friction.

Dry or ``Coulomb'' friction is a nonlinear dissipative force that
depends on the relative velocity $v$ between two sliding surfaces,
which, in the simplest model, takes the form $F_{frict} \sigma(v)$,
where $\sigma(v)$ is the sign of $v$ (and zero when $v=0$) and
$F_{frict}$ is the magnitude of the friction or contact force. The
effect of this force, as is well known, is to define a threshold
for external forces $F_{ext}$, such that if the surfaces are
relatively at rest ($v=0$), then the external force $F_{ext}$ does not induce motion
 unless $|F_{ext}| > |F_{frict}|$.  
This is the basis of stick-slip 
motion, which can be rendered all the more complex by the introduction
of time-dependent forces or external noises. P.-G. de Gennes~\cite{G05} showed,
in particular, that although a Brownian particle affected by dry
friction can never come to rest with ``pure'' Gaussian white noise,
its diffusive properties are very different from that of normal
Brownian motion with only viscous friction. In later
studies of de Gennes's and other models, it was also shown that
some features of stick-slip motion remain at the statistical
level, e.g., in the way in which the correlation function or the power
spectrum of Brownian motion shows a transition as a function of dry
friction and external forces; see \cite{TSJ10,BCT10,BTC11,TPJ12,BS12}.

Here we report on the experimental observation of some of
these properties relating to random motion with dry friction.  In our
experiment, sketched in Fig.~\ref{fig:setup}, a macroscopic rotator
has its axis suspended to a couple of bearings, which are the source
of dry friction, and is put in random motion by immersing it in a
fluidized granular gas~\cite{PL01}, which provides the noise source. The
motion of the rotator is recorded at high frequency, and its
statistical properties are analyzed within a suitable theoretical
framework, which includes a Boltzmann equation model, used in the
low collision regime, as well as a Langevin model in the high
collision regime. These two regimes are probed by changing the
fluidizing properties of the granular gas (e.g., the shaking
amplitude) and lead us to study two very different phenomenologies
associated with discontinuous (Poisson-type) noises, on the one hand, and
continuous (Gaussian-type) noises, on the other. Our results compare well with the
theoretical predictions obtained with the models mentioned above and provide a first experimental verification
of many of these predictions.  Overall, they also show that dry friction, which is
surely relevant in the discontinuous noise regime, has also an
important role in the continuous noise regime, even though the rotator
is kept in endless motion.

\section{Experimental setup}

\begin{figure}
\centering
\includegraphics[width=8cm,clip=true]{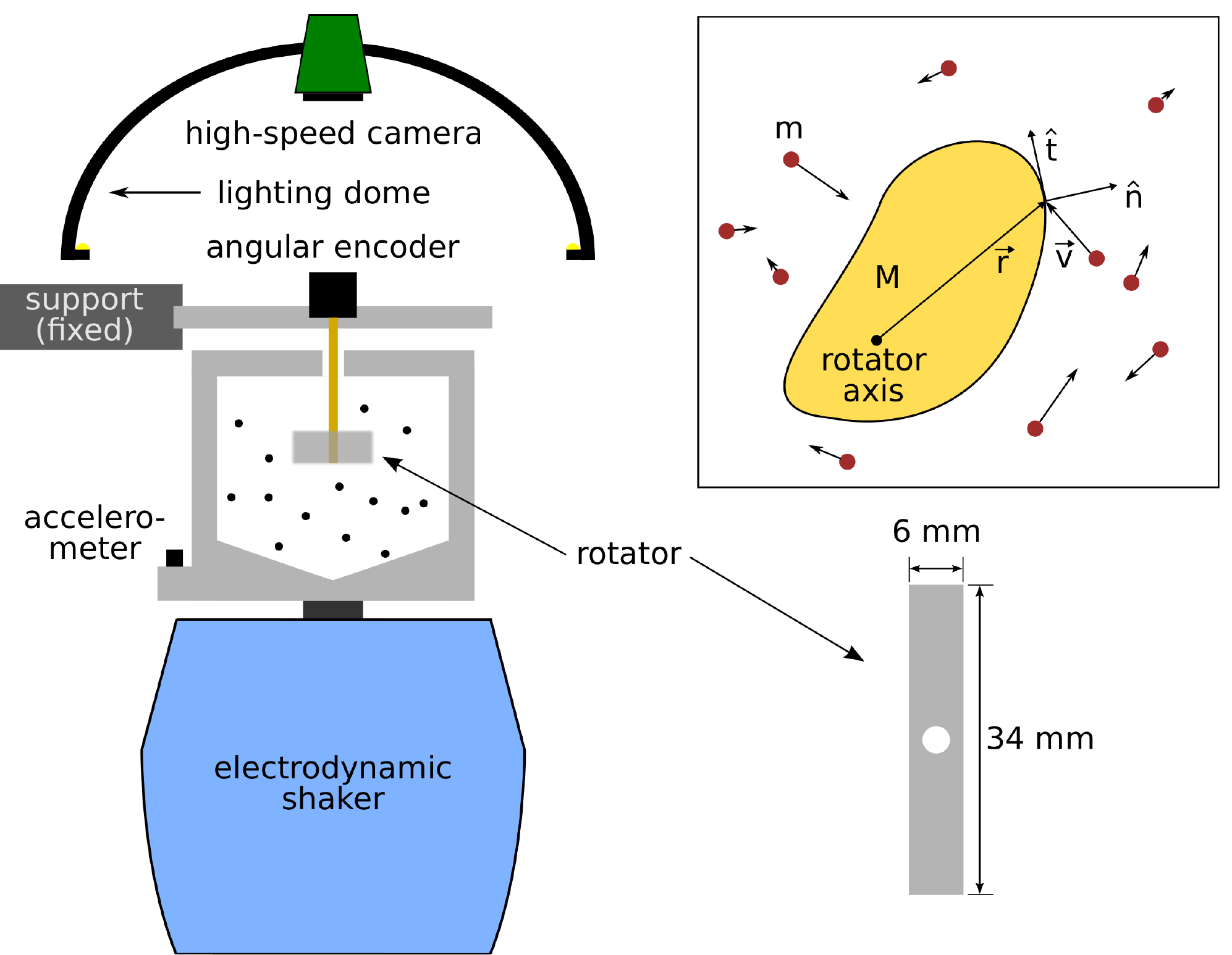}
\caption{Setup and definition of theoretical quantities.}
\label{fig:setup}
\end{figure}

The granular gas used is made of $N=50$ spheres of polyoxymethylene
(diameter $d=\SI{6}{\mm}$ and mass $m=\SI{0.15}{\gram}$) in a cylinder
of volume $V \approx \SI{1.9e5}{\mm^3}$ (and number density
$n=N/V$). The cylinder is shaken with a sinusoidal signal at
$\SI{53}{\hertz}$ and variable amplitude (measured by the maximum
rescaled acceleration $a_{\tmax}/g$ where $g$ is the gravity
acceleration). Suspended into the gas, a pawl (also called
``rotator'') of total surface $\Sigma=\SI{1.2e3}{\mm^2}$ (height
$h=\SI{15}{\mm}$ and base perimeter $S=\Sigma/h$), mass
$M=\SI{6.49}{\gram}$ and momentum of inertia $I=\SI{353}{\gram.\mm^2}$
rotates around a vertical axis attached to two ball bearings. The
position of the rotator is recorded in time by an angular
encoder (Avago Technologies). It is convenient to introduce the radius
of inertia $\ri=\sqrt{I/M}$ of the rotator. See Fig.~\ref{fig:setup}
for a sketch of the system and the definition of some quantities.

A close analysis of the dynamics of the rotator shows that it is well
described by the following equation of motion:
\begin{equation}
\dot{\omega}= -\Delta \sigma(\omega) -\gamma_a \omega + \eta_{coll}(t)
\end{equation}
where $\Delta=F_{frict}/{I}=\SI{38}\pm\SI{4}{\second^{-2}}$ is the
frictional force rescaled by inertia,
$\gamma_a=\SI{6}\pm\SI{1}{\second^{-1}}$ is some viscous damping rate
related perhaps to air or to other dissipations in the bearings, and
$\eta_{coll}(t)$ is the random force due to collisions with the
granular gas particles. The granular gas itself is stationary and
(roughly) homogeneous.

The velocity distribution of the spheres on the plane perpendicular to
the rotation axis is obtained by particle tracking via a fast camera
(see~\cite{PGGSV12} for details on the procedure) and is fairly
approximated by a Gaussian,
\begin{equation}
\phi(v) \sim e^{-v^2/(2v_0^2)},
\end{equation} 
where the
``thermal'' velocity $v_0$ has been introduced. Small deviations from
the Gaussian are observed but are neglected for the purpose of this
study; see~\cite{GPDGPSP12} for details. 
We have changed the maximum acceleration
rescaled by gravity $a_{\tmax}/g$ from $6$ to $20$, finding for
 $v_0$ values from
$\SI{200}{\mm^2.\second^{-2}}$ to $\SI{500}{\mm^2.\second^{-2}}$. 

The pawl is further characterized by its symmetric shape factor
$\langle g^2\rangle_{surf}=1.51$, where
$\langle \cdot \rangle_{surf}$ denotes a uniform average over the surface of
the object parallel to the rotation axis (see~\cite{CE08} for details).  The restitution coefficient
between the spheres and the pawl has been measured to be $\alpha \approx
0.83$. It is also useful to introduce
the ``equipartition'' angular velocity $\omega_0 = v_0 \eps/\ri$ where
$\eps=\sqrt{\frac{m}{M}}$. Note that, because of inelastic collisions
and frictional dissipations, the rotator {\em does not} satisfy
equipartition and $\omega_0$ is only a useful reference value.

\section{Boltzmann equation}

Since the packing fraction of the system does not exceed $3 \%$, the
single-particle probability density function (pdf) $p(\omega,t)$ of
the angular velocity of the rotator is expected to be fully described, under the
assumption of diluteness which guarantees molecular chaos, by the
following  equation~\cite{K61,CE08,TWV11}:
\begin{subequations} \label{beq}
\begin{align} 
\partial_t p(\omega,t)&=\partial_\omega [(\dat \sigma(\omega)+\gat
\omega)p(\omega,t)]+J[p,\phi] \\
J[p,\phi]&=\int d\omega'\, W(\omega|\omega')p(\omega',t)- p(\omega,t)
f_c(\omega),
\\
W(\omega'|\omega)&=\rho S \int \frac{ds}{S} \int d\bv\, \phi(\bv)
\Theta[(\bV(s)-
\bv) \cdot \hatn] \times \\ \nonumber &|(\bV(s)- \bv) \cdot
\hatn|\delta[\omega'-\omega-\Delta
\omega(s)],\\
\Delta \omega(s) &= (1+\alpha)\frac{[\bV(s)- \bv] \cdot
\hatn}{\ri}\frac{
g(s)\eps^2}{1+\eps^2g(s)^2}, \label{eq:colrule}
\end{align}
\end{subequations}
where we introduce the rates $W(\omega'|\omega)$ for the transition
$\omega \to \omega'$, the velocity-dependent collision frequency
$f_c(\omega)=\int d\omega'\, W(\omega'|\omega)$, the pdf $\phi(\bv)$ for the gas particle velocities, and
the so-called kinematic constraint in the form of Heaviside step
function $\Theta[(\bV- \bv) \cdot \hatn]$, which enforces the kinematic
condition necessary for impact. We also use the following symbols:
$\rho=n h$, ${\mathbf V}(s)=\omega \hat{z} \times {\mathbf r}(s)$ is the linear
velocity of the rotator at the point of impact ${\mathbf r}(s)$
parametrized by the curvilinear abscissa $s$ along the outer perimeter of
the rotator, $\hat{n}(s)$ is the unit vector perpendicular to the
surface at that point, and finally $g(s)={\mathbf r}(s)\cdot
\hat{t}(s)/{\ri}$ with $\hat{t}(s)=\hat{z} \times \hat{n}(s)$, which is
the unit vector tangent to the surface at the point of impact. We
refer to Fig.~\ref{fig:setup} for a visual explanation of these symbols.
The collision rule is given by
Eq.~\eqref{eq:colrule}~\cite{CE08}.  Note that in our setup, at homogeneous fluidization,
we measure $\rho S \approx \SI{0.31}{\mm^{-1}}$.

\section{Different regimes}

\begin{figure}
\centering
\includegraphics[width=8cm,clip=true]{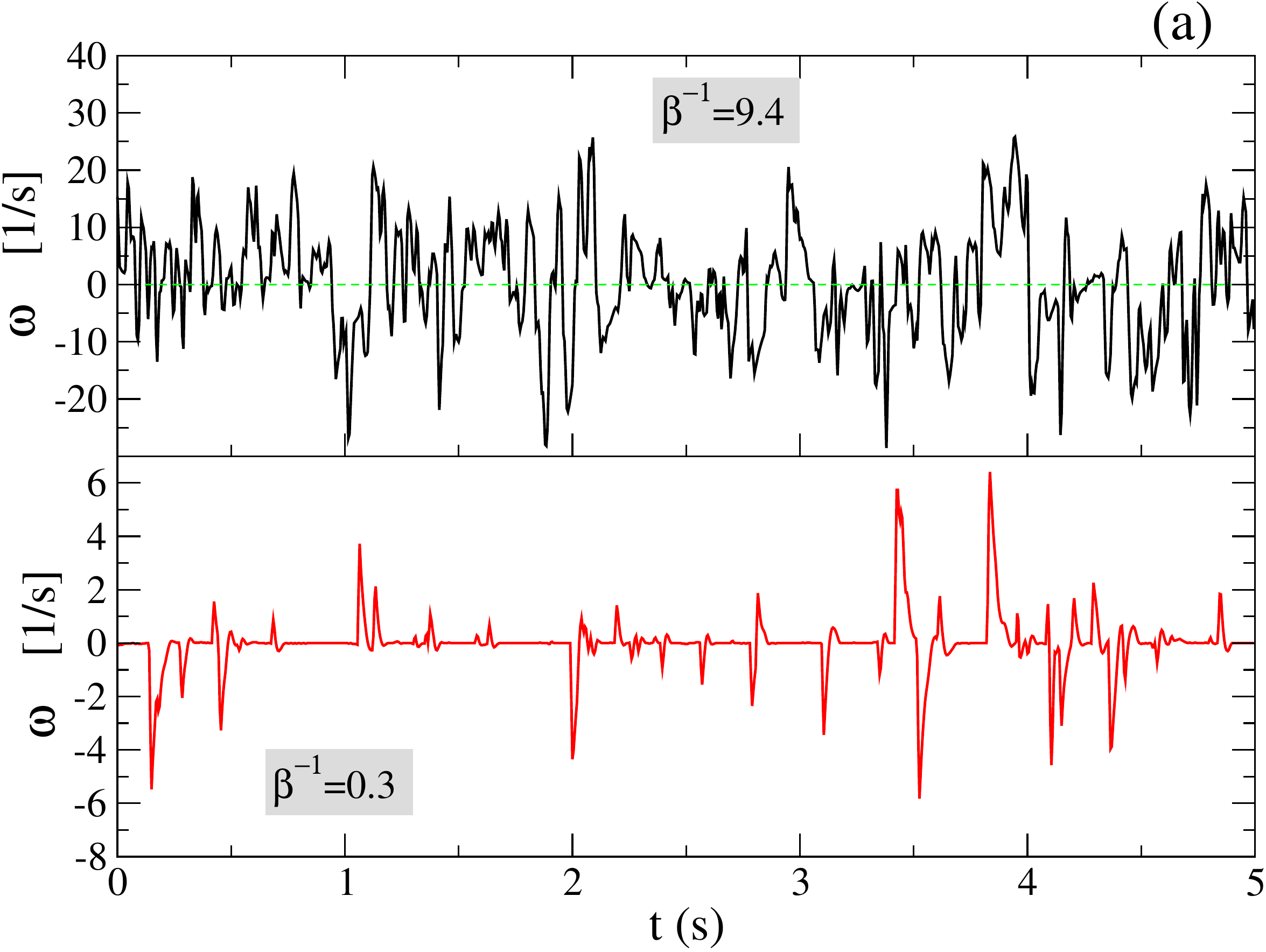}\\
\includegraphics[width=8cm,clip=true]{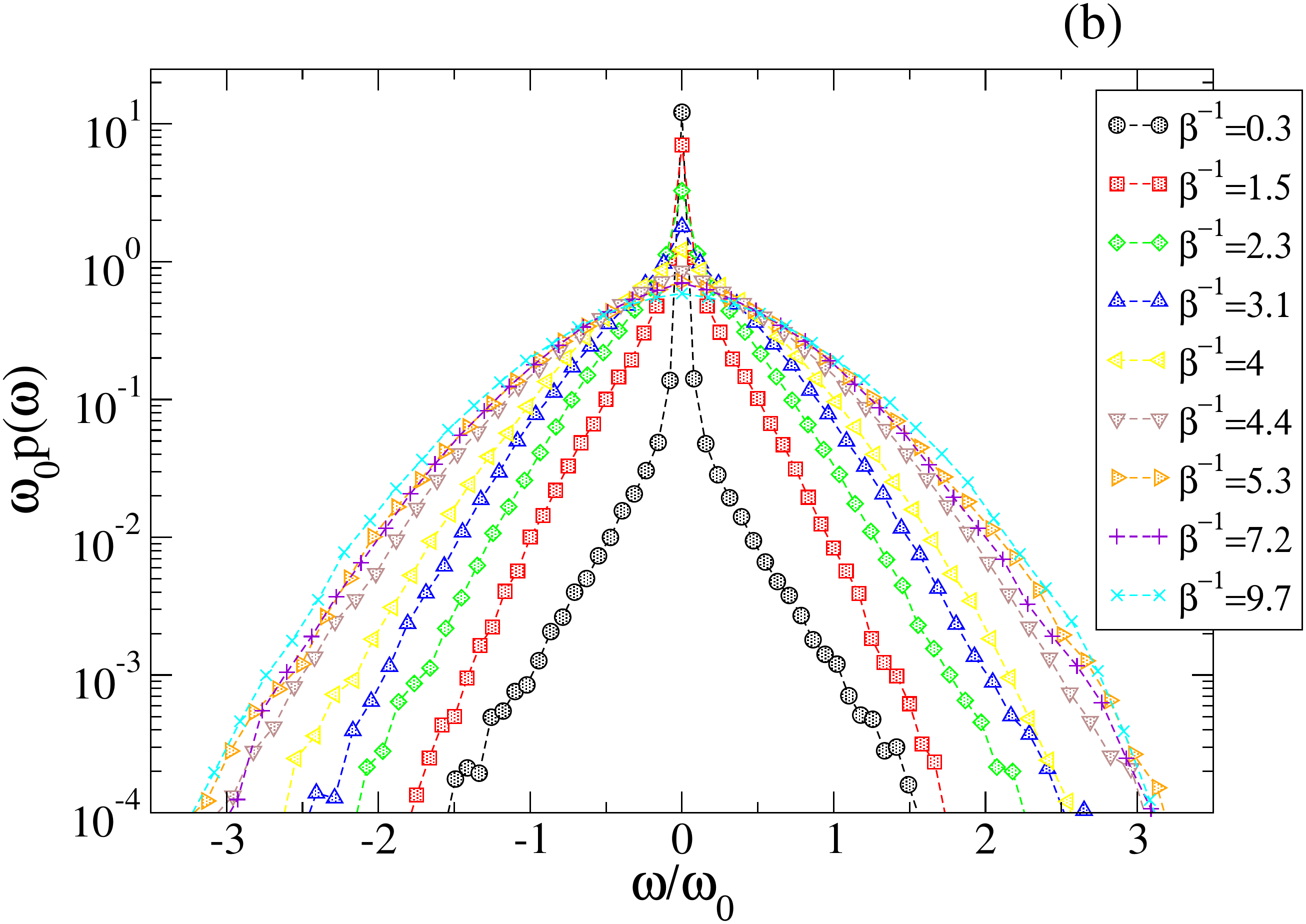}\\
\caption{(a) Two examples of signal $\omega(t)$ for different values of
  $\beta^{-1}$ in the experiment, corresponding to choices of the
  rescaled maximum acceleration $a_{\tmax}/g=4.1$ and $20.5$
  respectively; (b) rescaled experimental pdfs of the angular velocity for a range of rescaled
  accelerations going from $4.1$ to $21.2$. All other parameters are in the main text.\label{fig:pdfall}}
\end{figure}

An important parameter is
\begin{equation}
\beta^{-1} =\frac{\eps n\Sigma v_0^2}{\sqrt{2} \pi \ri \dat}
\approx\frac{\tau_\Delta}{\tau_c},
\end{equation}  
which estimates the ratio between the stopping time $\tau_\Delta \sim \omega_0/{ \dat}$ due to dissipation
(dominated by dry friction) 
and the collisional time $\tau_c \sim {(n \Sigma v_0)}^{-1}$.\footnote{Talbot \textit{et.\ al.}~\cite{TWV11} consider
a different parameter, namely, $\Gamma^*_s=\dat I/(\rho L^2 m v_0^2)$, and consider a
very thin rectangular rotator of length $L$ and a two-dimensional
projection of the system with density $\rho$, so that our $n\Sigma$ is
their $\rho (2 L)$, while our $\ri$ is their $L/(2\sqrt{3})$, leading to the
correspondence $\beta \to \sqrt{6} \pi \eps \Gamma^*_s$.}
A transition at $\beta \sim 1$ is
expected between a regime called the \textit{rare collision limit} (RCL) at
$\beta^{-1} \ll 1$, with the rotator at rest most of the time, and a
regime called the \textit{frequent collision limit} (FCL) at $\beta^{-1} \gg
1$, with the rotator always in motion, continuously perturbed by
collisions. The difference
between these two regimes is illustrated in Fig.~\ref{fig:pdfall}a.

The pdfs of the angular velocity obtained experimentally for values of $\beta^{-1}$
spanning the RCL and FCL
are reported in Fig.~\ref{fig:pdfall}b. There is a
great variability when $\beta^{-1}$ goes from small to
large values, i.e., when increasing the shaking amplitude and,
consequently, the collision frequency. At large values of
$\beta^{-1}$, the pdfs rescaled by $\omega_0$ tends to superimpose, a sign that
$\omega_0$ becomes the leading velocity scale. 
In order to make a more detailed contact with the theory and understand the
basic properties of the velocity pdf, we discuss next
the RCL and FCL regimes separately.

\section{Rare collision limit}

As seen above, the pawl in the RCL ($\beta^{-1} \ll 1$) is often
at rest, resulting in a peak around $\omega=0$ in the angular velocity
pdf. To describe this peak, we approximate the expected stationary pdf as
\begin{equation}
p(\omega)=a \delta(\omega) + (1-a)p_{smooth}(\omega),
\end{equation}
where $a$ is a suitable weight, decreasing as $\beta^{-1}$ grows, and $p_{smooth}(\omega)$ represents the smooth part of the pdf.

\begin{figure}
\centering
\includegraphics[width=8cm,clip=true]{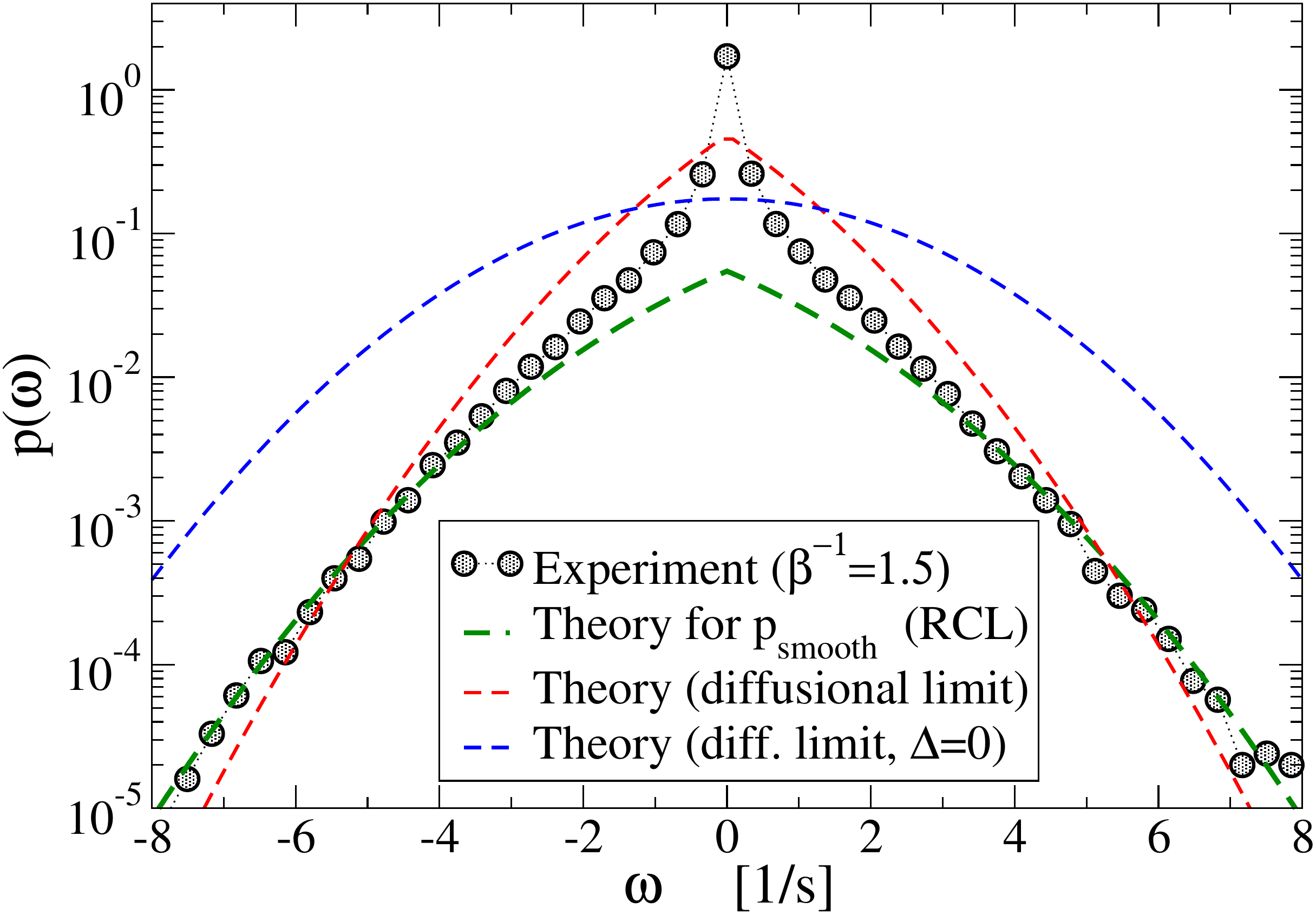}
\caption{Pdf of the pawl's angular velocity in the rare
  collisions limit (RCL), obtained with a maximum
  rescaled acceleration of the shaker given by $a_{\tmax}/g=6.5$ ($\beta^{-1}=1.5$). All other
  parameters of the experiment are given in the main text. The
  theoretical prediction~\eqref{eq:pdfrcl} is displayed as the dashed
  green line, where only $a$ is fitted with the experimental
  data. For reference, we also show the prediction of the theory
  in the diffusive limit with and without dry friction as the
  dashed red and blue curves, respectively. \label{fig:pdfrcl}}
\end{figure}

This form of stationary pdf has been studied
in~\cite{TWV11,BS12,baule2013}. In the RCL, the dynamics is reduced to
independent collisions followed by friction-induced relaxations. More
precisely, at a collision time $t$ the rotator velocity changes from
$0$ to $\omega^*$, depending on the projected impact velocity
$v=\mathbf{v}\cdot \hat{n}$ and the projected impact point
$g=\mathbf{r} \cdot \hat{t}/R_I$, and then relaxes according to
$\dot{\omega}=-\Delta \sigma(\omega^*)$ until a time $t+\tau$ such
that $\omega(t+\tau)=0$. In this case, the stationary average of any
function $y(\omega)$, restricted to the times where $\omega(t) \neq
0$, can be written as
\begin{equation}
\langle y \rangle = \rho S \int \frac{dS}{S} \int_{-\infty}^0 dv\, |v|
\phi(v)
\int_0^\tau dt\, y[\omega(t)].
\end{equation}
With this formula, we can calculate the characteristic function of $\omega$ by taking
$y=e^{i k \omega}$ and then invert the transform to retrieve
the smooth part $p_{smooth}(\omega)$. For the Gaussian $\phi(v)$ of variance $v_0^2$ and
the particular shape of our rotator, this yields
\begin{eqnarray}\label{eq:pdfrcl}
p_{smooth}(\omega) &=& \mathcal{N}'h\left(\frac{\omega}{(1+\alpha)\omega_0}\right),\\
h(x) &=&  \frac{e^{-2 x^2}}{4 |x|}
\left\{2-\textrm{erfc}\left[\frac{(\xi-1)|x|}{\sqrt{2\xi}}\right]
\right. - \nonumber\\
& &\quad\ \left. e^{2
x^2}\textrm{erfc}\left[\frac{(\xi+1)|x|}{\sqrt{2\xi}}\right]\right\}\\
\mathcal{N}'&=&\left[\int d\omega h\left(\frac{\omega}{(1+\alpha)\omega_0}\right) \right]^{-1}.
\end{eqnarray}

This result is compared with our experimental data in
Fig.~\ref{fig:pdfrcl}. It can be seen that the agreement of the tail
is very good, considering that we only fit the overall scaling factor $a$
representing the weight relative to the $\delta(\omega)$
contribution. On the contrary, the central part is not well
reproduced. We suspect that the discrepancy at low velocities is due
to a failure of the Coulomb friction model at those regimes. A
close inspection of single trajectories indeed reveals that the
free relaxing rotator frequently comes to rest with spurious
oscillations, likely to be due to the ball dynamics {\em
  inside} the bearings. This observation points to an interesting
application of studying ball bearings under random excitation:
in our case, the macroscopic observation of
$p(\omega)$ magnifies microscopic features around the zero velocity which would be hard to
characterize and understand otherwise.

From the experiment, we can also evaluate the autocorrelation $C(t)=\langle \omega(t)\omega(0) \rangle$ and power spectrum,
\begin{equation}
S(f) = \left|\sum_j \omega(t_j) e^{2 \pi \sqrt{-1} f t_j} \right|^2.
\end{equation}
These are shown in Fig.~\ref{fig:ps} as red curves. To our knowledge, no theory is
available for these quantities in the RCL. We notice that the large
frequency decay of the spectrum $S(f) \sim f^{-2}$ is compatible with
a small time exponential decay, while the part at small $f$ deviates
from it, suggesting a more rapid decay at large times. These features
are recovered in the graph of $C(t)$. Note that the power spectrum
also shows one of the higher harmonics of the shaker frequency
($3\times 53=\SI{159}{\hertz}$): it emerges only when the main signal due to the
dynamics of the rotator under the collisions becomes weak enough
and disappears in the FCL where the energy injected by
the collisions is larger.

\section{Frequent collisions and large rotator mass: equivalence with
continuous
white noise}

In the FCL, it is useful to exploit the difference of mass (here $\eps=\sqrt{m/M}=0.15$) by taking a further $\eps \ll 1$ limit. Such
a limit is often called a ``diffusive limit'' and allows us to expand the
Boltzmann equation~(\ref{beq}) to obtain a Fokker-Planck equation or,
equivalently, a Langevin equation for the pawl
velocity~\cite{CE08} having the form
\begin{equation} \label{lang2}
\dot{\omega}=-\gamma \omega -\Delta \sigma(\omega) +
\sqrt{\Gamma_g}\,\xi,
\end{equation}
where $\gamma=\gamma_a+\gamma_g$, $\gamma_g$ represents a granular viscosity, $\gamma_a$ an air
viscosity, $\Gamma_g$ a granular velocity diffusion coefficient, and $\xi$ is a Gaussian white noise
with unit variance. In our setting, $\gamma_g$ and
$\Gamma_g$
are given by~\cite{CE08}
\begin{subequations}\label{eq:coeff}
\begin{align} 
\gamma_g &=
(1+\alpha)\sqrt{\frac{2}{\pi}}\rho S\frac{m}{M}v_0\langle g^2
\rangle_{surf}\\
\Gamma_g &= (1+\alpha)\gamma_g \frac{m}{I} v_0^2.
\end{align}
\end{subequations}

The study of the Langevin equation~\eqref{lang2} was initiated in~\cite{CD61},
received strong impulse by de Gennes in~\cite{G05}, and was completed
in~\cite{TSJ10,BCT10,BTC11,TPJ12}. Its stationary velocity
distribution reads \cite{H05}
\begin{align} \label{eq:pdf}
p(\omega)&= \mathcal{N}
\exp\left[-\frac{(|\omega|+\Delta/\gamma)^2}{\Gamma_g/\gamma}\right],\\
\mathcal{N}^{-1}&=\sqrt{\pi\Gamma_g/\gamma}\;\textrm{erfc}(\Delta/\sqrt
{\gamma
\Gamma_g}).
\end{align}
 Note that in the limit $\beta \to 0$, e.g., when dry friction disappears ($\Delta \to 0$), the
 stationary pdf goes to a Gaussian of variance $\Gamma_g/\gamma$.
Moreover,
 assuming also $\gamma_a \ll \gamma_g$ (which is consistent with the
FCL), one has $\Gamma_g/\gamma \approx \omega_0^2 (1+\alpha)/{2}$, so
that
 equipartition with the gas $\langle \omega^2 \rangle = \omega_0^2$ is
satisfied
in the ideal elastic case
 $\alpha=1$~\cite{SVCP10}.

\begin{figure}
\centering
\includegraphics[width=8cm,clip=true]{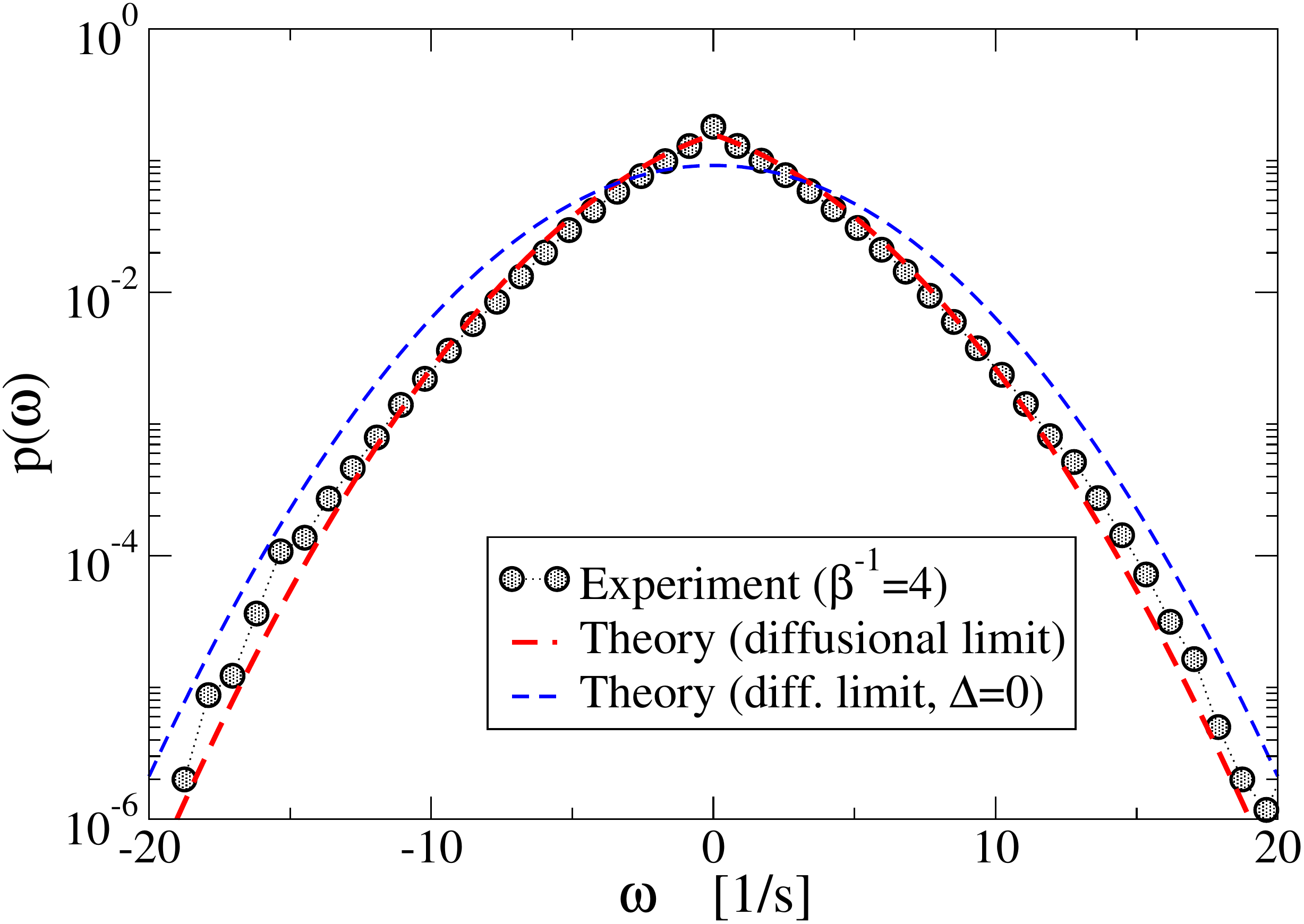}
\caption{Pdf of the pawl's angular velocity in the frequent
  collisions regime (FCL) obtained when the maximum rescaled
  acceleration of the shaker is $a_{\tmax}/g=11.74$. The
  prediction of the diffusive limit theory, Eq.~\eqref{eq:pdf},
  is displayed as the dashed red line with {\em no fitting
    parameters}. For reference, we also show with the dashed blue line
  the prediction of the diffusive limit without dry friction. \label{fig:pdfdiff}}
\end{figure}

In Fig.~\ref{fig:pdfdiff}, we find good agreement between the pdf above and the experimental
data. This comparison, obtained with {\em no fitting parameters},
represents one of the first known experimental verification of
the velocity pdf~\eqref{eq:pdf}, as well as one of the few
experimental applications of the diffusive limit of granular kinetic
theory. The cusp at $\omega=0$ predicted by the theory is a striking
consequence of the presence of Coulomb friction, and is well reproduced in
the experimental data. At large velocities, $\omega \gg \Delta/\gamma
\approx \SI{6}{\second^{-1}}$, the pdf recovers Gaussian tails due to the dominance
of linear viscosity.

\begin{figure}
\centering
\includegraphics[width=8cm,clip=true]{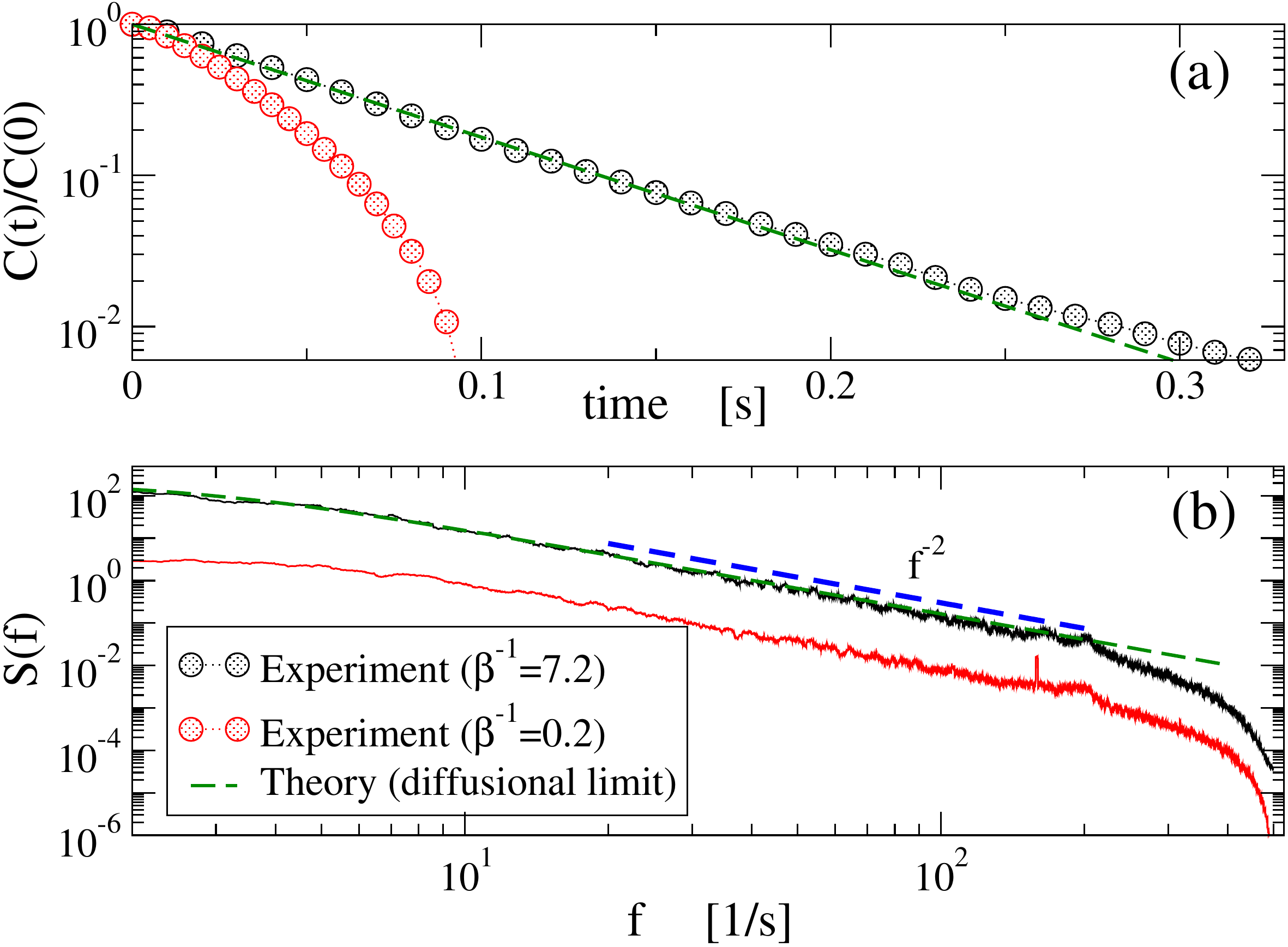}
\caption{(a) Autocorrelation and (b) power spectrum of the pawl's angular
velocity in the FCL (black) and RCL (red) obtained with a maximum rescaled
acceleration of the shaker corresponding to $a_{\tmax}/g=4.1$ and $17$, respectively.  
The diffusive limit prediction is superimposed as the dashed green
line.\label{fig:ps} }
\end{figure}

In the diffusive limit, where Eq.~\eqref{lang2} holds, theoretical
expectations for the autocorrelation of the angular velocity and for
the power spectrum have been obtained analytically in~\cite{TSJ10,TPJ12}. These
expressions involve the eigenvalues of the Fokker-Planck operator, and are too long to
be reproduced here.
A verification of these expressions, shown in Fig.~\ref{fig:ps}, 
indicates that the Langevin model of Eq.~\eqref{lang2} offers a good description of our
experiment in the FCL. 

\section{Relevance of friction in the FCL}

\begin{figure}
\centering
\includegraphics[width=7.7cm,clip=true]{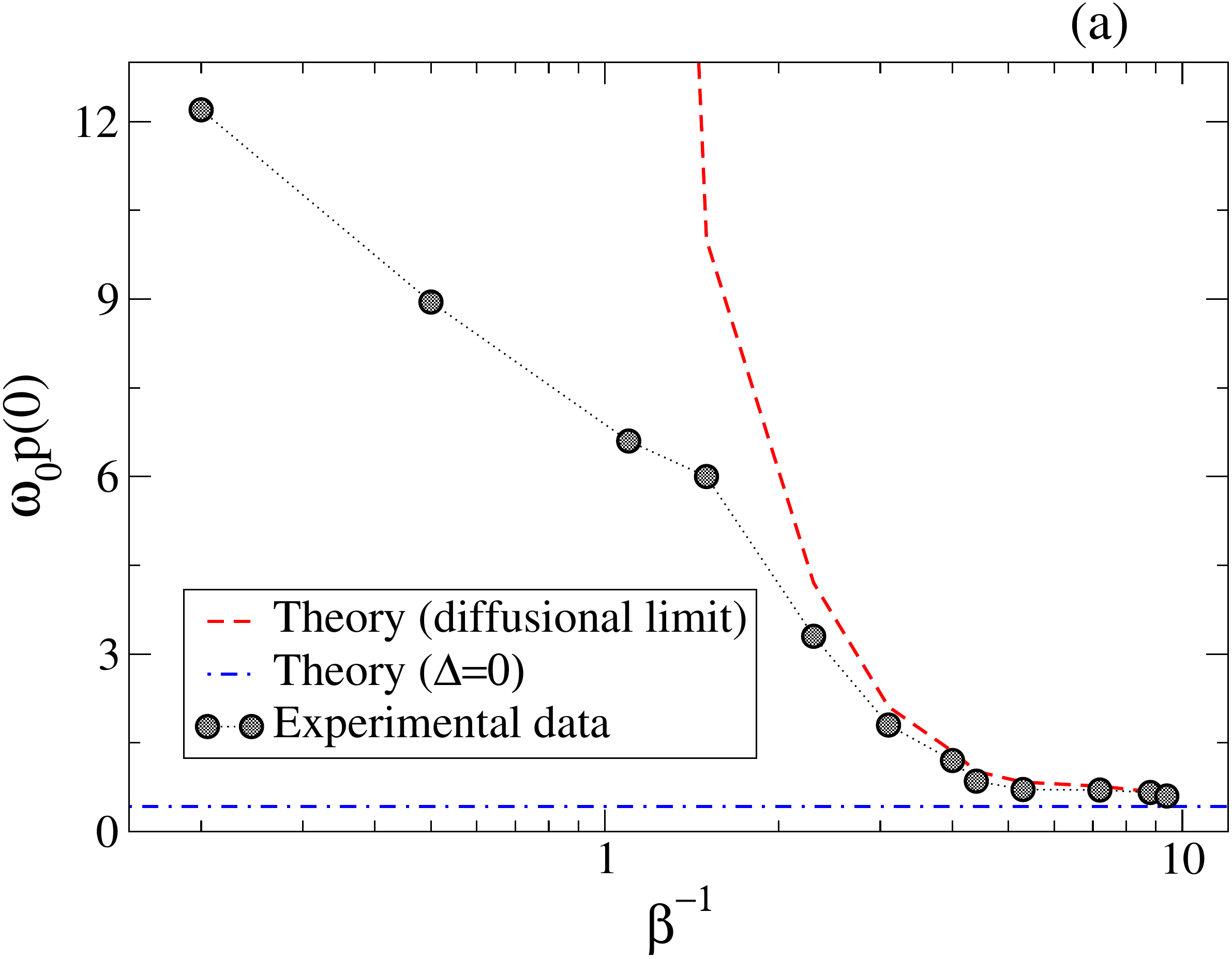}\\
\includegraphics[width=7.9cm,clip=true]{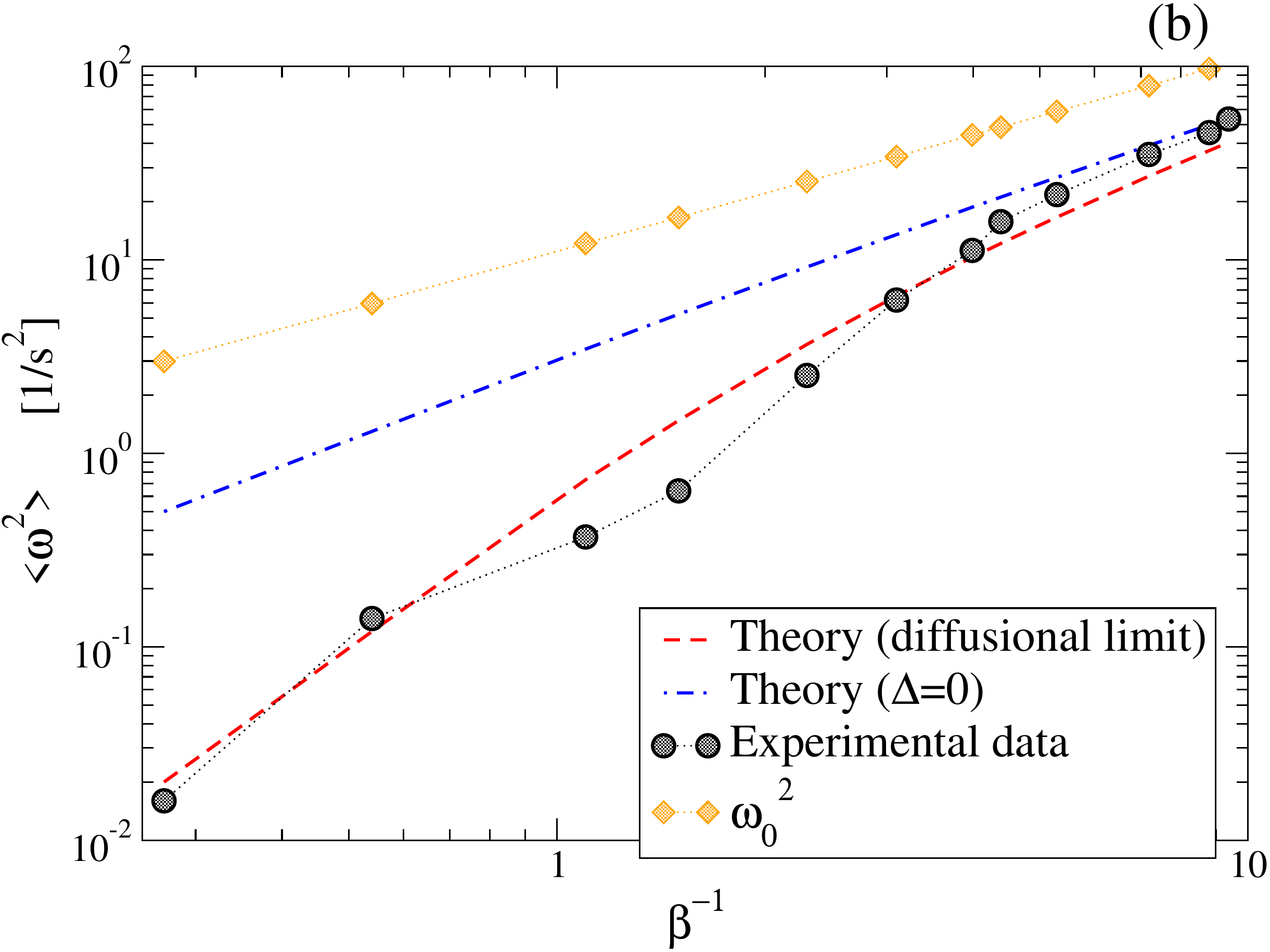}\\
\includegraphics[width=7.9cm,clip=true]{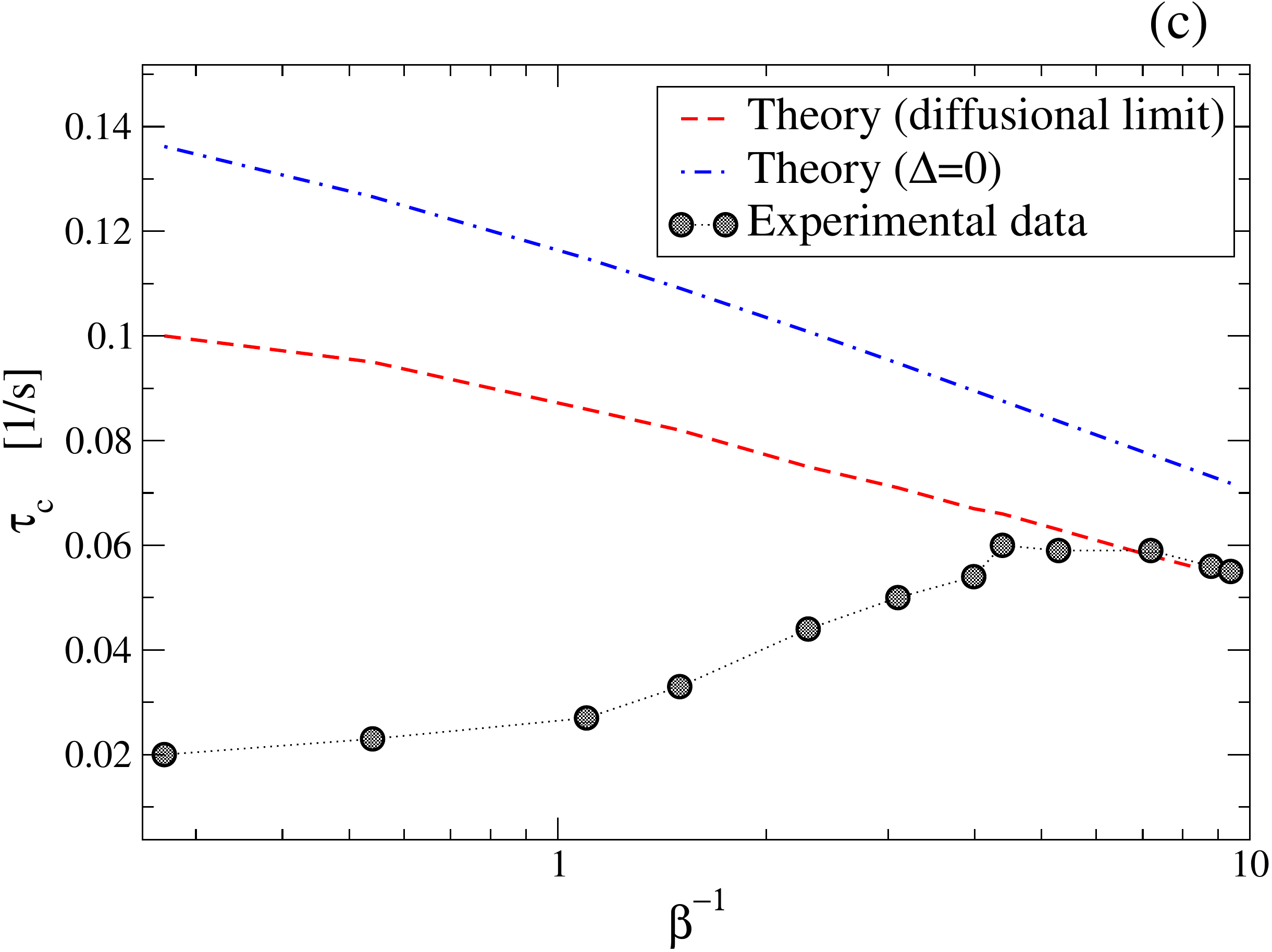}
\caption{Comparison of experiments (black points) \textit{versus} theory in the
diffusive
limit with dry friction (red lines) and without dry friction (blue lines),
  in order to evaluate the relevance of friction as
$\beta^{-1}$
is varied: (a) rescaled peak of the velocity pdf; (b)
  variance of the velocity pdf; (c) decay time of the velocity
autocorrelation.
\label{fig:regimes}}
\end{figure}

A more refined comparison of experiment and theory is obtained in
Fig.~\ref{fig:regimes}, where three main observables are plotted
against $\beta^{-1}$: the rescaled peak of the velocity distribution,
the variance of the distribution, and the correlation time $\tau_c$
obtained by fitting $C(t) \sim \exp(-t/\tau_c)$. The three figures
also show the predictions, for each choice of the parameters, of the diffusive limit 
theory, as given by Eq.~(\ref{lang2}),
together with the predictions of the same theory in the absence of dry friction
($\Delta=0$). This last comparison is useful to evaluate the relevance
of the dry friction term. 

In Fig.~\ref{fig:regimes}a, we display the
peak of the experimental pdf (again rescaled by $\omega_0$) as a function of
$\beta^{-1}$. This information, in the RCL when the rotator is most of
the time at rest, is an indirect probe of $a$, since the experimental value
$p(0)$ includes also $(1-a)p_{smooth}(0)$. This figure clearly shows
the decrease of the peak as $\beta^{-1}$ increases. Moreover, it shows
that, when $\beta^{-1} > 1$, such a peak gets closer to the values
analytically computed in the diffusive limit where $\Delta$ is negligible
and the pdf tends to a Gaussian. 
Interestingly, the variance of the distribution $\langle \omega^2\rangle$, whose
formula in the diffusive limit reads
\begin{equation} \label{eq:var}
\langle \omega^2\rangle = \left(\frac{\Delta}{\gamma}\right)^2 +
\frac{\Gamma_g}{2 \gamma} - \mathcal{N} \frac{\Delta
\Gamma_g}{\gamma^2}
\exp\left(-\frac{\Delta^2}{\gamma \Gamma_g}\right),
\end{equation}
adheres to the prediction of the diffusive theory even in the RCL,
while at large $\beta^{-1}$ both the experiment and the diffusive theory go
toward the linear limit, where $\Delta$ is negligible. The small
discrepancies are likely to be due to the finiteness of the mass ratio $\eps$ in the
experiment.  

A different scenario is observed for the
correlation time $\tau_c$, as the experiment shows a non-monotonous
behavior when plotted against $\beta^{-1}$, with $\tau_c$ {\em
  growing} when moving from the RCL to the FCL, up to a regime where
the experiment is well described by the diffusive theory (red
line). This regime displays a slight decrease of $\tau_c$ as
$\beta^{-1}$ is further increased. It is interesting to note that,
even at the highest values of $\beta^{-1}$, the prediction of the
linear theory without friction ($\Delta=0$) overestimates by roughly
$40\%$ the experimental results. The correlation time is therefore
very sensitive to the presence of the dry friction and signals
it even when the ``static'' information coming from the velocity pdf
is basically not affected.

\section{Conclusions}

We have discussed in this letter the results of an experiment in which a rotator is
submitted to dry friction and collisions with a granular gas. By
tuning the shaking amplitude at constant frequency, we have explored
 different random, Brownian-like dynamics which are either 
dominated by friction or by collisions, as well as the crossover
between these two extreme regimes. In the rare collision regime (RCL), our 
data for the 
velocity pdf display a macroscopic fraction of events at rest
($\omega=0$) and non-Gaussian tails at high velocity, which are both well
reproduced by a Boltzmann collision model. In the frequent
collision regime (FCL), our results for the velocity pdf, autocorrelation and power
spectrum are well explained by a Langevin model with dry and viscous frictions~\cite{TSJ10}, 
which can be derived from the Boltzmann model in the diffusive limit.
In this limit, dry friction tends to become
negligible compared to the other forces; this explains why, at very high collision frequency, one recovers a
phenomenology partly explained by the Ornstein-Uhlenbeck model, if we exclude the correlation decay. 

To conclude, we remark that our experiment suggests a useful way to
estimate parameters which are somewhat difficult to measure directly,
in analogy with Einstein's theory of Brownian motion which gives
access to Avogadro's number through a macroscopic measurement. In our
case, Eq.~\eqref{eq:var} together with the expressions shown in
\eqref{eq:coeff} may be used to obtain estimates of $\alpha$ or
$\Delta$ knowing the other parameters and the value of $\langle
\omega^2 \rangle$.  At the same time, our experiment offers a positive
test of the ability of kinetic theory to predict macroscopic Brownian
coefficients, such as the viscosity $\gamma_g$ and the noise amplitude
$\Gamma_g$. Ongoing extensions of our study include the coupling of
the system with a motor~\cite{N12} in order to apply an external force
and investigate its interplay with friction and collisions
\cite{TSJ10}, and the possibility of observing ratchet effects for
asymmetric rotators~\cite{TWV11,joubaud2012,GPDGPSP12}.

\newpage
\acknowledgements

The authors acknowledge the support of the
Italian MIUR under the grants FIRB-IDEAS no.\ RBID08Z9JE, 
FIRB no.\ RBFR081IUK and no.\ RBFR08M3P4, and PRIN  no.\ 2009PYYZM5.

\bibliographystyle{unsrt}
\bibliography{fluct}

\end{document}